\documentclass[a4paper]{article}

\catcode`\é=13  \defé{\'e}
\catcode`\â=13  \defâ{\^a}
\catcode`\è=13  \defè{\`e}
\catcode`\à=13  \defà{\`a}
\catcode`\ù=13  \defù{\`u}
\catcode`\ç=13  \defç{\c c}
\catcode`\ê=13  \defê{\^e}
\catcode`\ô=13  \defô{\^o}
\catcode`\û=13  \defû{\^u}
\catcode`\ù=13  \defù{\`u}
\catcode`\î=13  \defî{\^\i}
\catcode`\ï=13  \defï{\"\i}

\usepackage{fullpage} 
\usepackage{pstricks}
\usepackage{amssymb}
\usepackage{amsthm}
\usepackage{amsmath} 
\usepackage{amsfonts}
\usepackage[version=3]{mhchem}
\usepackage{multicol}
\usepackage{hyperref}
\usepackage{graphicx}
\usepackage{color}
\usepackage{soul}
\usepackage[margin=1.7cm,top=3cm,bottom=2.8cm,centering]{geometry}
\usepackage[english,francais]{babel}

\begin{document}
\selectlanguage{english}

\title{A model of dynamic stability of H3K9me3 heterochromatin to explain the resistance to reprogramming of differentiated cells \footnote{Reference: Jehanno C, Flouriot G, Le Goff P and Michel D. 2017. A model of dynamic stability of H3K9me3 heterochromatin to explain the resistance to reprogramming of differentiated cells. Biochim Biophys Acta. 1860, 184-195. doi: 10.1016/j.bbagrm.2016.11.006.}}
\author{Charly Jehanno$ ^{a} $, Gilles Flouriot$ ^{a} $, Pascale Le Goff$ ^{a} $ and Denis Michel$ ^{a} $ \footnote{denis.michel@live.fr}\\
\newline
$ ^{a} $\begin{small}Universite de Rennes1-IRSET. Campus sant\'e de Villejean 35000 Rennes, France. \end{small}}
 
\date{}
\maketitle
\begin{multicols}{2}

Despite their dynamic nature, certain chromatin marks must be maintained over the long term. This is particulary true for histone 3 lysine 9 (H3K9) trimethylation, that is involved in the maintenance of healthy differentiated cellular states by preventing inappropriate gene expression, and has been recently identified as the most efficient barrier to cellular reprogramming in nuclear transfer experiments. We propose that the capacity of the enzymes SUV39H1/2 to rebind to a minor fraction of their products, either directly or via HP1$ \alpha/\beta $, contributes to the solidity of this mark through (i) a positive feedback involved in its establishment by the mutual enforcement of H3K9me3 and SUV39H1/2 and then (ii) a negative feedback sufficient to strongly stabilize H3K9me3 heterochromatin in post-mitotic cells by generating local enzyme concentrations capable of counteracting transient bursts of demethylation. This model does not require direct molecular interactions with adjacent nucleosomes and is favoured by a series of additional mechanisms including (i) the protection of chromatin-bound SUV39H1/2 from the turnovers of soluble proteins, which can explain the uncoupling between the cellular contents in SUV39H1 mRNA and protein; (ii) the cooperative dependence on the local density of the H3K9me3 of HP1$ \alpha/\beta $-dependent heterochomatin condensation and, dispensably (iii) restricted enzyme exchanges with chromocenters confining the reactive bursts of SUV39H1/2 in heterochromatin. This mechanism illustrates how seemingly static epigenetic states can be firmly maintained by dynamic and reversible modifications.

\section{Introduction}

\subsection{H3K9me3 and differentiation}
The resistance and inheritance of repressive chromatin modifications are supposed to result from complex networks of interactions between specific DNA regions and chromatin-modifying enzymes, mediated by a plethora of molecular actors including non coding RNA, co-repressors and methyl-cytosine binding proteins linking DNA methylation to histone modifications. However, a simpler picture recently emerged when showing that modifications forced at arbitrary DNA regions can be propagated as well by histone modifying enzymes over cellular generations \cite{Ragunathan}. The number of actors necessary to stabilize epigenetic states will also be limited in the simplified and original model presented here for the mark H3K9me3, that is crucial for preserving cellular integrity \cite{Peters} and forbidding dedifferentiation. Indeed, this mark has recently been shown remarkably resistant and identified as the primary barrier preventing reprogramming in nucleus transfer experiments \cite{Matoba,Antony}. The transplanted nuclei carry in a purely epigenetic manner the factors ensuring the stable maintenance of the cellular differentiation status. H3K9me3 appears as the security bolt for silencing the genes incompatible with this differentiation, a role previously foreseen for DNA methylation supposedly less dynamic. In fact, a loss of H3K9me3 but not of DNA methylation, was observed in the model of chromatin loosening of WRN-deficient cells \cite{Zhang} and earlier studies showed that DNA methylation follows H3K9 trimethylation \cite{Lehnertz,Feldman}. \\
H3K9me3 strongly increases during normal cell differentiation \cite{Hawkins} and conversely, low H3K9me3 and diffuse euchromatin characterize stem cells \cite{Meshorer,Kraushaar}, consistent with their background of generalized gene expression \cite{Efroni} and their self-renewal \cite{Loh}. SUV39H1 and SUV39H2 are the major enzymes responsible for H3K9 trimethylation, because their inactivation in embryonic cells is not compensated by other enzymes with the same activity such as SETDB1 \cite{Peters1}.\\
The reverse process of chromatin loosening has been reported in cancer cells which acquire globally open chromatin with prominent histone acetylation and lower H3K9me3 \cite{Meshorer,McDonald,Flouriot}. The loss of heterochromatin in cancer is also reflected by the decondensation of Barr bodies (X inactivated) in many breast and ovarian cancers \cite{Pageau}. The weakening of constitutive heterochromatin causes genome instability \cite{Slee} while that of facultative heterochromatin leads to deregulated gene expression, with possible reactivation of tumor promoting genes. A drop of H3K9me3/SUV39H1 heterochromatin is also clearly associated with aging \cite{Zhang}. The H3K9 trimethylase SUV39H1 has been shown anti-tumorigenic \cite{Braig,Peters,Khanal} whereas conversely, JMJD2 enzymes which have the opposite activity, have been shown oncogenic \cite{Liu,Slee}. Various H3K9me3 demethylases are induced during hypoxia \cite{Beyer}. Upregulation of JMJD2A/KDM4A by gene amplification or hypoxia, leads to gene copy gains through re-replication \cite{Whetstine}. Certain reports on H3K9me3 in cancer are however conflicting since an increase of H3K9me3 and SUV39H1 activity has also been observed in malignant cells. This seeming paradox could in fact have been expected if considering the contractile phenotype of metastatic cells as a secondary differentiation state different from that of the cell at the origin of the cancer, for instance epithelial \cite{Jehanno}. 

\subsection{Heterochromatin organization}

Two main types of silencing hetereochromatin (associated to either H3K9me3 or H3K27me3 marks) are differently used by vertebrate species \cite{Gilbert} and may be involved in different circumstances in a given species. H3K9me3 has been proposed to precede long term polycomb-mediated chromatin closure \cite{Bernstein} but inversely, H3K27me3 has also been suggested to be a temporary repression signal controlling developmental genes \cite{Kim}. These two marks can coexist and concern different subsets of chromosomal regions. H3K27me3  is mainly found in regions including unmethylated CpG-rich sequences, whereas H3K9me3 regions generally contain methylated DNA. H3K9me3 heterochromatin appears malleable \cite{Hathaway}, and has stabilizing and repressive roles for constitutive heterochromatin (repetitive DNA and centromeres) and facultative heterochromatin (developmentally turned off genes). 
H3K9me3 heterochromatin is specifically bound to heterochromatin protein 1$ \alpha/\beta $ (thereafter collectively written HP1), a protein involved in its condensation into densely packed structures known as chromocenters, historically characterized in cytology by their capacity to trap more dye (such as Feulgen or methylene blue), and now often called pericentromeric heterochromatin (PCH). The nuclear organization of the chromocenters is strangely variable. It is perinuclear and perinucleolar in MCF7 human cells but take the form of well-delimited spheres scattered throughout the nucleus in murine cells \cite{Almouzni}. As this organization greatly facilitates experimental studies, murine cells like 3T3 mouse cells are generally selected, what will be done here also. Interestingly in nocturnal mammals, heterochromatin otherwise located at the nuclear periphery, accumulates at the center of the nucleus of photoreceptor cells, probably for convenience to facilitate light sensing \cite{Solovei}. These disparate locations tend to minimize the importance of nuclear architecture to regulate heterochromatin. Although chromocenters are mainly associated to constitutive heterochromatin, they can also include facultative heterochromatin. Genes irreversibly silenced during development \cite{Hubner}, LINE retroposons \cite{Bulut} and E2F-target genes in post-mitotic differentiating cells \cite{Guasconi} can also relocate into these structures. Interestingly, the presence of SUV39H1 bound to H3K9me3 has been evidenced by chromatin immunoprecipitation in LINE, before their progressively decrease and replacement by DNA methylation \cite{Bulut}. The formation of H3K9me3 heterochromatin depends on the density of H3K9me3 \cite{Hamada} but not on DNA sequences, as clearly illustrated by the lack of sequence conservation of centromeric DNA  between species \cite{Almouzni}.

\subsection{Dynamic stability}

Cells and their epigenetic marks are open systems with permanent molecular renewal, in which no molecular complexes can be definitely locked \cite{Kamakaka,memories}. Chromatin modifications were initially considered as static epigenetic marks, but have then been shown highly reversible and labile, continuously written and erased by antagonistic chromatin-modifying and -demodifying enzymes present together in the cell \cite{Huang}. Nevertheless, adult somatic cells must securely maintain their differentiation status in long-lived species. This apparent problem is solved by mechanisms called dynamic stability.
	For example, in differentiated cells, H3K9me3/HP1$ \alpha/\beta $ heterochromatin is stable for years in spite of the very fast exchange of HP1, initiating the concept of dynamic stability \cite{Cheutin}. The rapid turnovers of chromatin modifications imposed us to look for new, dynamic mechanisms ensuring the steady-state stability of chromatin configurations. Following the pioneer model of \cite{Dodd}, positive feedbacks mediated by various interactions between histone-interacting and sometimes DNA-interacting machineries, have been proposed to ensure the copying of chromatin modifications during DNA replication and the regional spreading of chromatin states \cite{Probst,Moazed,Zhu}. The role of certain chromatin-modifying enzymes, including SUV39H1, in this mechanism has been attributed to their capacity to rebind to their enzymatic product and to the existence of a rate of cis-propagation of enzymatic reactions from modified sites to adjacent unmodified sites in the vicinity \cite{Hodge,David-Rus,Hathaway,Muller}. Mechanisms called nucleation and looping have recently been developed \cite{Rippe,Erdel}, involving in the case of SUV39H, complex interaction networks including HP1 and methyl-DNA binding proteins \cite{Rippe}. In this latter model, an immobile fraction of SUV39H catalyses the formation of H3K9me3 in the surrounding region, whose extent is restricted by the possibility of DNA looping. But even in absence of replication copying and following their initial establishment, chromatin modifications should be maintained for years in post-mitotic differentiated cells, that is to say during time windows much larger than the turnovers of all chromatin-modification and molecular immobility. New mechanisms are proposed here, which are based on the dynamic rebinding of SUV39H enzymes to its products and can contribute to the solidity of H3K9me3 heterochromatin and by this way prevent cancer and increase longevity. The phenomenon of rebinding of chromatin-modifying enzymes, that is not specific to SUV39H, is generally assumed to underly the propagation and mitotic memory of chromatin marks through information transfers between modified and unmodified nucleosomes. In the present model, the rebinding of SUV39H1/2 to their products participates to the robustness of established heterochromatin. This mechanism is original in that (i) it does not rely on ternary complexes and direct physical transfers between modified and unmodified nucleosomes, (ii) it is functional at a regional scale, for example in chromocenters, and not restricted to adjacent nucleosomes and (iii) it is capable of ensuring the resistance of chromatin marks once established, by preparing, in a latent form, a massive source of free SUV39H capable of counteracting a stimulus responsible for its release. 

\section{Experimental observations}

The model proposed below is built on reports from the literature and a few additional observations. The ideal material for studying the maintenance of H3K9me3 in adult cells would be post-mitotic tissular cells, but as they can not be easily cultured and manipulated, cell lines are generally used instead. These cell lines are however interesting in that they differ in their index of epithelial dedifferentiation, as reflected for instance by the degree of disruption of E-cadherin adhesion complexes \cite{Flouriot}. Fig.1 shows the comparison of four widely used cell lines, two retaining a relatively high differentiation status: the mammalian MCF7 mammary epithelial cells and the human HepG2 hepathocytic cell line; and two with a more de-differentiated status: the human MDA-MB-231 and Hela cells. Compared to these two latter ones, the first two ones are characterized by higher levels of H3K9me3 and H3K9 trimethylase SUV39H1, and lower levels of H3K9ac and H3K9me3 demethylase JMJD2C/KDM4C. These measurements further support the association between global H3K9me3 and differentiation. 

\begin{center}
\includegraphics[width=5.5cm]{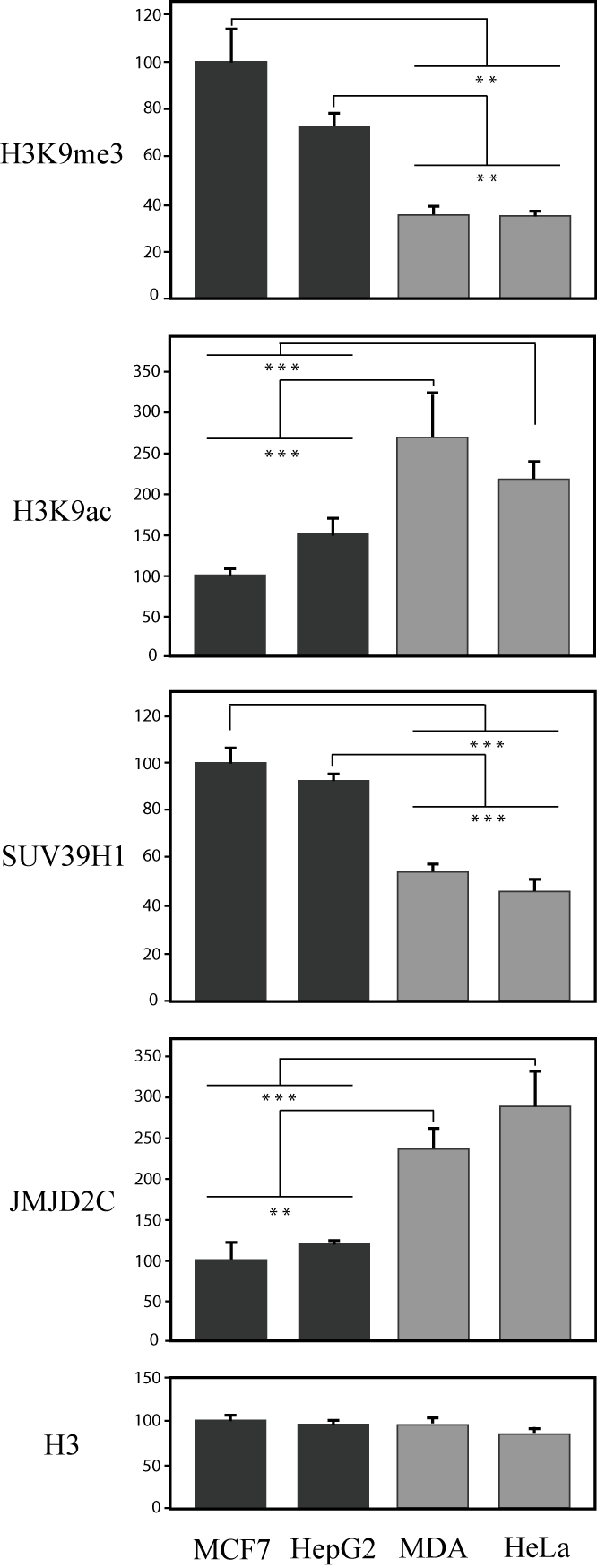} \\
\end{center}
\begin{small} \textbf{Figure 1} Comparison of H3K9 trimethylation and acetylation and SUV39H1 and JMJD2C proteins in human cell lines known for their low (MCF7 and HepG2) or high (MDA-MB-231 and Hela) dedifferentation grade. The fluorescence intensity was quantified and normalized by the mean intensity obtained in MCF7 cells arbitrarily set to 100. Anova one-way test was used. Asterisks indicate significant differences between each cell line for a given quantification (* P $ < $ 0.05; ** P $ < $ 0.01; *** P $ < $ 0.001). \end{small}\\

\subsection{Regulation of SUV39H1 with the degree of heterochromatinization}

To minimize the number of parameters, instead of comparing different cells, it is preferable to study heterochromatin in the same cell type. We developed a cellular tool to decrease heterochromatin in the human MCF7 cells, using the stable expression of a mutant version of MKL1 involved in mechano-signalling, deleted from its N-terminal region ($ \Delta $N200) \cite{Flouriot}. Heterochromatin loosening triggered by stable expression of this deletion mutant of MKL1 using a tetracyclin-inducible expression system, significantly decreases H3K9me3. The mechanism of heterochromatin disruption caused by MKL1-$ \Delta $N200 is currently unknown, but good candidates for mediating this effect are the JMJD2 demethylases which are upregulated. Interestingly, an important decrease of the SUV39H1 protein, but not of the SUV39H1 mRNA, was observed, by western blot (Fig.2A) and by cyto-immunofluorescence (60\% decrease), compared to control MCF7 cells stably transfected with the empty vector.

\begin{center}
\includegraphics[width=7cm]{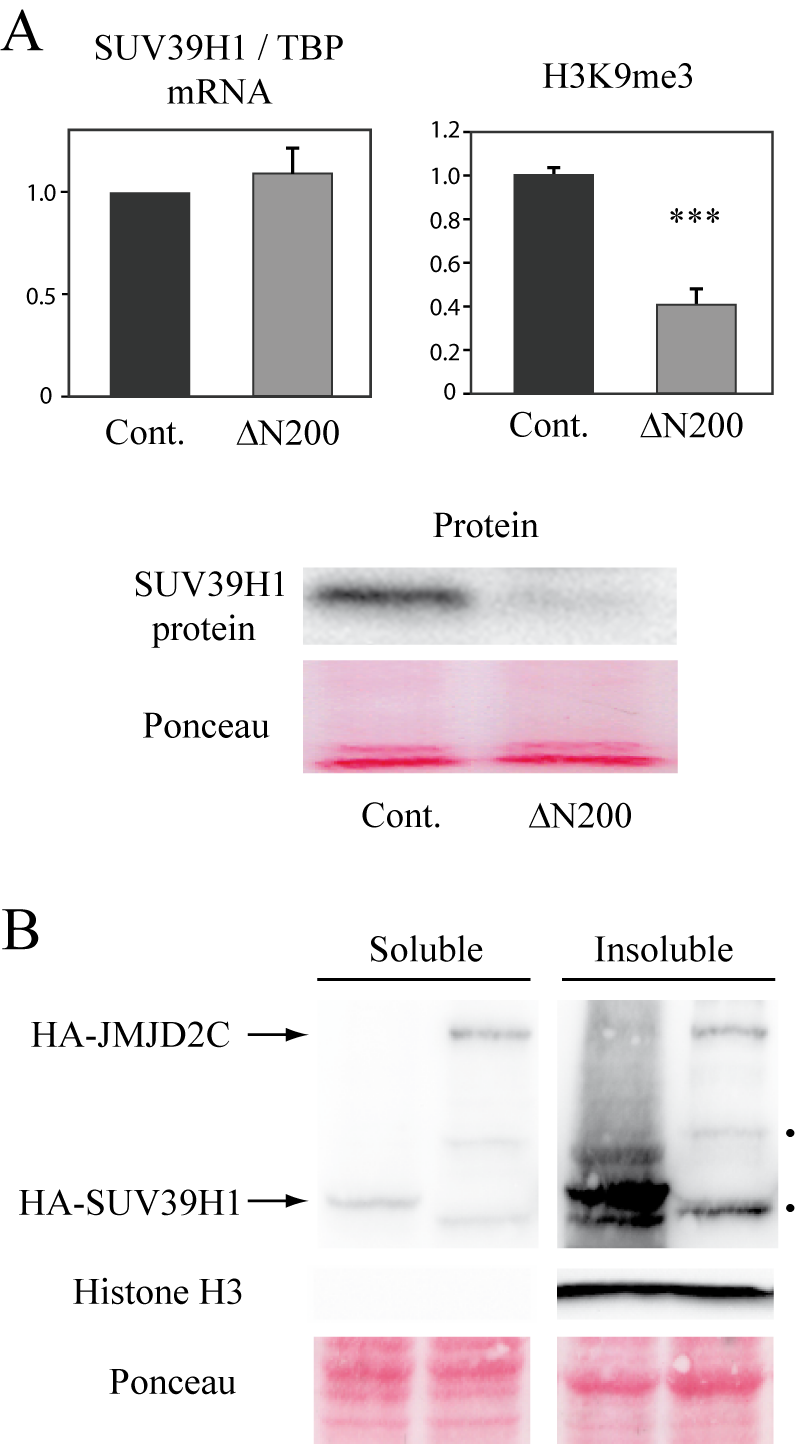} \\
\end{center}
\begin{small} \textbf{Figure 2} SUV39H1 mRNA and protein in the cell. (\textbf{A}) Down regulation of the SUV39H1 protein in H3K9me3-poor in MCF7 $ \Delta $N200 cell line, but not of the SUV39h1 mRNA. The fluorescence intensity of the H3K9me3 mark was performed as previously described using ImageJ. Mann-Whitney test was used. Asterisks indicate significant differences between each condition, control and $ \Delta $N200 (*** P $ < $ 0.001). (\textbf{B}) Soluble and insoluble fractions of MCF7 cells transiently expressing HA-tagged SUV39H1 or JMJD2C, were probed by immunoblotting using the same antibody directed against the HA epitope. Contrary to JMJD2C, SUV39H1 strongly accumulates in the insoluble fraction containing chromatin ($ \bullet $ nonspecific signals).\end{small}\\

 On the one hand, the presence of high levels of H3K9me3 in cells containing more SUV39H1 is obviously expected since H3K9me3 is the product of the SUV39H1 enzyme. But on the other hand, the origin of the high amount of SUV39H1 in cells containing more H3K9me3 should be explained. As shown in Fig.2, it can not result from a higher SUV39H1 gene expression level because the mRNA content is unchanged. Alternatively, the fact that SUV39H1 binds to H3K9me3 and a series of observations from the literature, suggest the possibility that SUV39H1 protein can be stabilized by fixation on chromatin. This post-transcriptional mechanism would provide an explanation for the apparent disconnection between the regulation of the SUV39H mRNA and protein (Fig.2). Indeed, a decrease of the SUV39H1 protein has been shown during heterochromatin loosening in earlier studies, which reported that the phosphorylation of Serine 391 of SUV39H1 induces simultaneously the dissociation of SUV39H1 from chromatin \cite{Park} and its proteasomal destabilization \cite{Khanal}. This dual effect of the modification of the same residue shows that SUV39H1 solubilization and degradation are correlated events, and that the insolubilization of SUV39H1 in heterochromatin could protect it from degradation. Under this hypothesis, SUV39H1 is expected to be less stable in absence of H3K9me3 heterochromatin, for example following a pulse of JMJD2A and massive H3K9me3 demethylation which has been shown to solubilize SUV39H1 \cite{Black}. This is a cycle of reciprocal influences between H3K9me3 and SUV39H1, in which SUV39H1 increases the level of H3K9me3 and conversely H3K9me3 stabilizes the fraction of SUV39H1 to which it is bound. Biochemical fractionation experiments using the control cells transfected with either HA-tagged SUV39H1 and HA-tagged JMJD2C, showed that the large majority of HA-SUV39H1 is present in the insoluble fraction which contains all the chromatin fraction as verified using H3 immunoreactivity (Fig.2B).

\subsection{SUV39H1 is reversibly bound to chromatin}
Most SUV39H1 appears bound to chromatin (Fig.2B), but fluorescence recovery after photobleaching (FRAP) analyses showed that it is in fact subject to permanent association-dissociation cycles, with different mean cycling times ranging from seconds to minutes depending on the fractions \cite{Krouwels,Rippe}. In self-renewing cells expressing SUV39H1-GFP, the recovery of fluorescence in bleached areas shows that all the SUV39H1 fractions are interchangeable. This reversibility of the association between SUV39H1 and chromatin is in line with the removal of SUV39H1 caused by demethylation and acetylation of H3K9. For example, the mobility of SUV39H1 is increased by experimental expression of JMJD2/KDM4 enzymes \cite{Black,Rippe}.\\
 
\subsection{Chromocenter formation and SUV39H1-HP1 functional relationships}

The role of HP1 in the rebinding of SUV39H1 to H3K9me3 is unclear. The enzymatic activity of SUV39H1 has been shown necessary for its accumulation at heterochromatic sites \cite{Lachner,Bannister}. The accumulation of SUV39H on chromatin is clearly a rebinding phenomenon, not related to its catalytic activity, because the enrichment of SUV39H1 in H3K9me3-dense regions is also observed for a catalytic mutant of SUV39H1 (Fig.S1B). This accumulation was considered indirect and mediated by HP1 \cite{Nakayama,Haldar}. But SUV39H1 has also been shown self-sufficient for H3K9me3 binding in vitro \cite{Wang} and in vivo \cite{Krouwels,Melcher,Muller}, which is expected in principle since SUV39H1/2, like HP1, have a chromodomain. The partial disruption of heterochromatin by hyperacetylation using Trichostatin A (TSA) treatment, leads to the complete dislocation of HP1 from the residual chromocenters, whereas SUV39H1 still binds them \cite{Rippe}. This latter result suggests that HP1 is, at least partially, dispensable for SUV39H1 binding to established heterochromatin. The proper enzymatic activity and subcellular localization of SUV39H1 are not altered by fusion with a fluorescent protein irrespective of whether it is linked to its C-terminus or its N-terminus, corresponding to its domain of interaction with HP1, suggesting that it is not critical for chromatin binding (data not shown). The accumulation of HP1 and SUV39H1 in heterochromatin could proceed through different mechanisms \cite{Krouwels,Muller}, which do not forbid interactions between these molecules. In this study, we will not assume a particular hypothesis on the involvement of HP1 on SUV39H binding. SUV39H will be supposed to rebind to H3K9me3 either directly or indirectly through HP1. HP1 will be nevertheless essential for its established role in the condensation of H3K9me3 heterochromatin. The binding of HP1 to H3K9me3 and its condensation-promoting role are cooperative. Through its capacity to cross-link nucleosomes, HP1 triggers the condensation of H3K9me3-rich regions \cite{Canzio,Teif,Mulligan,Azzaz}. Such a mechanism has not been reported for SUV39H1/2. HP1 will be assumed to condense DNA in a nonlinear manner over a critical density of H3K9me3 marks. The density of H3K9me3 has been suggested to be only slightly higher inside than outside the chromocenters \cite{Rippe}. If chromocenters are defined by the intensity of DAPI staining, the intensity of the H3K9me3 signal appears much higher in these areas compared to the rest of the nucleoplasm (top panels of Fig.S1A), but this sharp difference can result from two causes difficult to distinguish: the proportion of H3K9me3 and the degree of compaction. However, the near complete exclusion of H3K9ac labelling from these condensed regions (Fig.S1A) suggests that the different marks are significantly segregated in and out of the chromocenters.

\section{General principles applying to chromocenters}

\subsection{Ligand rebinding to receptor arrays}

The large arrays of H3K9me3 moities of the chromocenters to which diffusing SUV39H1 can bind, is a typical case of clustered receptors. The capacity of clustered receptors to bias ligand rebinding has been abundantly investigated, mainly for cell surface receptors \cite{Care,DeLisi,Goldstein1989,Goldstein1995,Gopalakrishnan,Thompson}, but without clear consensus. Both association and dissociation rates have been suggested to be affected by high receptor density. These studies introduced several notions, like (i) the "encounter complexes" \cite{DeLisi,Shoup}, which are virtual complexes in which the ligand is neither free nor bound but close enough to bind, or (ii) a parameter named likelihood of prompt rebinding ($ b $) defined when ligand rebinding is favoured in areas containing clustered receptors \cite{Thompson}. By contrast, other authors calculated that receptor clustering should inhibit ligand rebinding compared to scattered receptors, without modifying the microscopic rate constants \cite{Care}. Here, a simpler mechanism only based on mass action rules will be retained. Receptor clustering will be supposed to not modify the absolute binding and dissociation rates, but to influence the local concentration of free ligand in two ways. 
(\textbf{i}) The receptor clusters (H3K9me3 arrays) carried by heterochomatin is a crystal-like insoluble structure which can save the bound ligand from the normal turnover of soluble proteins continuously synthesized and degraded. By this way, the arrays of H3K9me3 can ensure the protection of SUV39H1 by insolubilizing it. (\textbf{ii}) As long as it is sequestered by its reaction product, SUV39H1 is enzymatically inactive, but it nevertheless constitutes a stock of enzyme which can be resolubilized in case of disappearance of H3K9me3 sites, for example following a rise in concentration of demethylating enzymes JMJD2, as observed in \cite{Black}. Altogether, these observations allow to rationally design a simple model to explain the resistance of H3K9me3 heterochromatin.

\section{The modeling tools}

The probabilities of the different nucleosome states are given by sets of ordinary differential equations (ODEs). As the number of nucleosomes is large, stochastic approaches like the master equation are not necessary. A particularity of this model is a permutation of relative concentrations depending on the scale considered. The following deterministic treatment mixes single molecule and bulk approaches to analyse the reciprocal influences between individual molecules and the population of molecules. For a given nucleosome in a given gene, the histone-modifying enzymes appear numerous and can be incorporated into pseudo-first order binding constants. Conversely, because of enzyme rebinding to a fraction of H3K9me3 residues, the concentration of enzymes depends on the huge number of nucleosomes present in the nucleus. The concentration of nucleosomes should now take part to pseudo-first order constants for determining the amount of active enzymes by multiplying the probability of single nucleosome state by the number of nucleosomes. The probability of individual nucleosome state corresponds to the proportion of nucleosomes in this state in the vicinity, and the concentrations of diffusing enzymes determined at the macroscopic level can be incorporated into pseudo-first-order constants in the single molecule treatments. The sequestration of the enzymes by the substrates will be neglected during the enzymatic reactions (large Michaelis constants $ K_{M} $) to not consider zero-order ultrasensitivity mechanisms and better focus on the present model. SUV39H1 is considered as a processive enzyme which has to rebind to the histone after each round of methylation \cite{Chin}. As the geometry of the nucleosolic and heterochromatin compartments is unknown, sophisticated diffusion schemes will be avoided. The concentration of S-adenosyl-methionine will be considered as non-limiting and the nucleo-cytoplasmic shuttling of SUV39H1 will be considered passive. 

\section{Formal treatment}

\subsection{The actors and nomenclature}
Several enzymes have the same H3K9me3-demethylase activity. JMJD2C has been shown involved in the self-renewal of embryonic stem cells \cite{Loh}, but JMJD2B has also been proposed to antagonize pericentric heterochromatin \cite{Fodor} and JMJD2A has been shown to massively demethylate H3K9me3 \cite{Black}. As a matter of fact, different enzymes can be used equivalently to facilitate the experimental reprogramming in single cell nuclear transfer experiments, like JMJD2D \cite{Matoba} and JMJD2B \cite{Antony}. Given the variety of JMJD2 enzymes intervening in the different cellular contexts, they will be collectively named JMJD2. In the same manner, as SUV39H1 and SUV39H2 appear interchangeable \cite{Rice}, they will be gathered under the generic name SUV39H.\\
Although the residue H3K9 exists in many different states, only three will be considered here. Each nucleosome is supposed to switch between 3 possible states:  H3K9-dimethylated ($ N_{a} $), H3K9-trimethylated ($ N_{b} $) and H3K9-trimethylated bound to SUV39H ($ N_{b}S $) (Fig.3). For simplicity, the other states of H3K9: monomethylated, unmethylated, or acetylated, are supposed, in a first approximation, to remain in constant ratio relatively to dimethylated H3K9.

\subsection{Nonlinear dependence on H3K9me3 of heterochromatin condensation}

Heterochromatin condensation depends on the density of H3K9me3 and on HP1, without need for complex hierarchical regulations \cite{Hamada}. Several studies pointed the cooperative mechanisms of HP1-mediated clustering of chromatin regions containing H3K9me3 marks \cite{Canzio,Teif,Mulligan}. Accordingly, the opposite mark H3K9ac appears excluded from the chromocenters (Fig.S1A). The nucleosome-bridging activity of rapidly moving HP1 tends to bring together regions with high H3K9me3 density and in turn, the higher concentration of H3K9me3 further facilitates HP1 recruitment, thus yielding a self-reinforced loop leading to the coalescence of heterochromatin into chromocenters. This positive feed back underlies the apparent cooperativity of HP1-mediated condensation which introduces a nonlinear dependence on the density of H3K9me3. As a consequence, a decrease of the concentration of H3K9me3 below a certain threshold can induce a dramatic dismantlement of chromocenters. For modeling this nonlinear effect, a theoretical heterochromatin condensation index ($ CI $) will be defined to evaluate the effects of H3K9me3 variations on condensation, using a steep Hill function

\begin{equation} CI = \dfrac{(N_{b}+N_{b}S)^{\bar n}}{\left(K\frac{N}{HP1}\right )^{\bar n}+(N_{b}+N_{b}S)^{\bar n}} \end{equation}
where $ N_{b}+N_{b}S $ is the total concentration of H3K9me3, considering that HP1 can bind as well to H3K9me3 in presence and absence of SUV39H. $ N $ is the total concentration of  H3K9, HP1 is the concentration, supposed large and fixed, of HP1$ \alpha $ or $ \beta $ molecules. $ K $ is a dissociation constant and $ \bar n $ is a Hill coefficient. $ \bar n $ and $ K $ can be such that faint modifications of H3K9me3 density can have major effects on condensation. This index will allow to visualize the critical dependence of condensation on the stability of H3K9me3 (Figs.4-6).

\subsection{Enzymes concentrations}

For easier treatment, the proportions of these forms will be used $ x =[N_{a}]/N, y =[N_{b}]/N, z =[N_{b}S]/N $, with $ N = [N_{a}]+[N_{b}]+[N_{b}S] $

The schemes with and without SUV39H rebinding are represented in Fig.3. 

\begin{center}
\includegraphics[width=5cm]{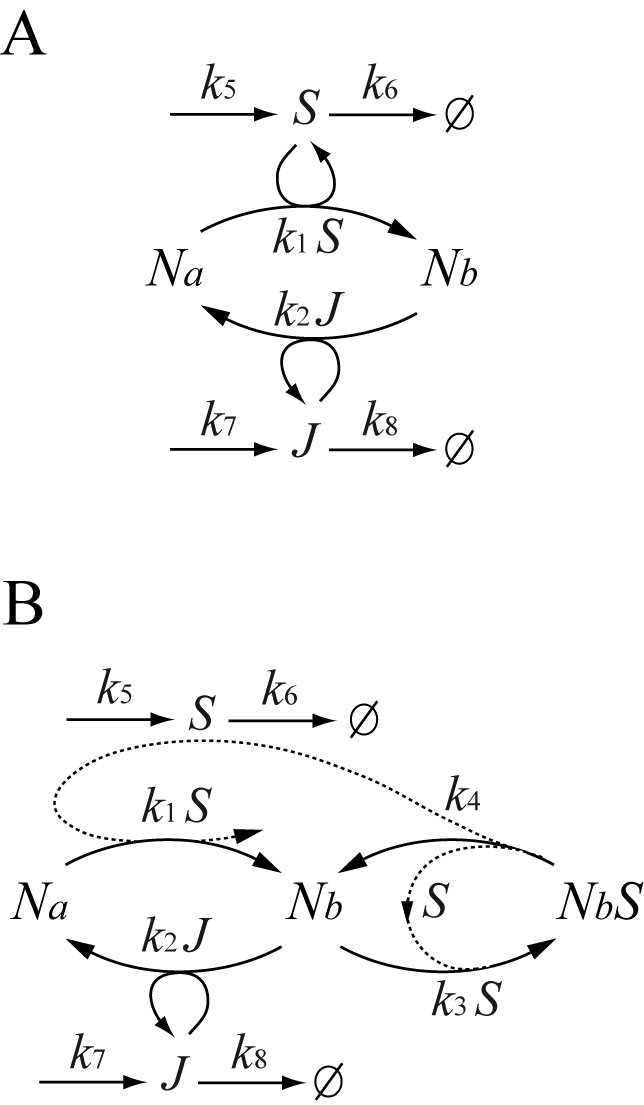} \\
\end{center}
\begin{small} \textbf{Figure 3}. Simplified models of trimethylation/detrimethylation of H3K9, (\textbf{A}) without enzyme sequestration, (\textbf{B}) with enzyme sequestration. A single nucleosome is supposed for simplicity to exist in only 3 forms: H3K9me1/2 ($ N_{a}$), H3K9me3  ($ N_{b} $) or H3K9me3/SUV ($ N_{b}S $), with probabilities $ x, y,$ and $ z $ respectively. $ S $: SUV39H, $ J $: JMJD2. \end{small}\\ 

\subsection{Without SUV39H rebinding to its product (Fig.3A)}

\begin{equation} \dot{x}= k_{2}J y-k_{1}S x \end{equation}
with $ y=1-x $. The initial steady state conditions are simply $ x= k_{2}J/(k_{1}S+k_{2}J) $, $ y= k_{1}S/(k_{1}S+k_{2}J) $, where the steady state levels of free enzymes are $ S=k_{5}/k_{6} $ and $ J=k_{7}/k_{8} $. Let us now suppose that $ S $ can rebind to its product $ N_{b} $.

\subsection{SUV39H rebinding to its product (Fig.3B)}

Now the dynamic system can be described as follows:

\begin{subequations} \label{E:gp}
\begin{equation} \dot{x}= k_{2}J y-k_{1}S x \end{equation} \label{E:gp1}
\begin{equation} \dot{y}= k_{1}S x + k_{4}z-y (k_{2}J +k_{3}S ) \end{equation} \label{E:gp2}
\begin{equation} \dot{z}= k_{3}S y-k_{4} z \end{equation} \label{E:gp3}

The concentration of the free enzymes depends on the concentrations of the relative populations of nucleosomes. These concentrations can be simply obtained by multiplying the concentration of nucleosomes $ N $ by their state probabilities.

\begin{equation} \dot{S}= k_{5}+k_{4}Nz-S (k_{6} +k_{3}Ny) \end{equation} \label{E:gp4}
\begin{equation} \dot{J}= k_{7}-k_{8}J \end{equation} \label{E:gp5}
\end{subequations} 

$ x $,  $ y $ and $ z $ are probabilities ($ x+y+z=1 $), whereas $ S $, $ J $ and $ N $ are concentrations. The switch from probabilistic to quantitative ODEs is obtained by multiplying molecule state probabilities by the concentrations of this molecule. The initial steady state conditions are 

\begin{subequations} \label{E:gp}
\begin{equation} x_{0}= k_{4}k_{2}J_{0} /D \end{equation} \label{E:gp1}
\begin{equation} y_{0}= k_{4}k_{1}S_{0} /D \end{equation} \label{E:gp2}
\begin{equation} z_{0}= k_{1}k_{3}S_{0}^{2} /D \end{equation} \label{E:gp3}
where \textit{D} is the sum of the numerators of $ x_{0}, y_{0}, z_{0} $
and the concentrations of free enzymes are

\begin{equation}  S_{0}=\dfrac{k_{5}}{k_{6}} \ \ J_{0}=\dfrac{k_{7}}{k_{8}} \end{equation} \label{E:gp4}
\end{subequations} 

\subsection{Perturbative test of these systems}

The systems described above are monostable (i.e. have a unique steady state fixed by the relative values of the kinetic constants), but they can have interesting stabilizing effects. The influence of SUV39H rebinding on the resistance to changes can be evaluated using a sudden perturbation like a burst of JMJD2, shifted from $ J_{0} $ to $ J_{1} $ while maintaining $ S_{0} $ constant. This perturbation is biologically relevant because JMJD2 has been shown to have the capacity to dislodge SUV39H from chromatin \cite{Black} and it is upregulated during hypoxia \cite{Beyer,Whetstine}, in cancer \cite{Liu,Slee} and in association with dedifferentiation (Fig.1).

\subsubsection{With SUV39H rebinding}
Condensation can resist a tenfold increase of JMJD2 (Fig.4). By accumulating SUV39H and by protecting it from the continuous degradation of soluble proteins, the sequestration of SUV39H by heterochromatin generates a stock of enzyme mobilisable in case of demethylation. A peak of JMJD2 demethylases triggers an outburst of soluble SUV39H that can in turn strongly counteract demethylation, even when keeping constant the rates of synthesis and degradation of SUV39H. For this effect, there is no need for H3K9me3 to be the predominant mark, nor that all H3K9me3 moieties are occupied by SUV39H.

\subsubsection{Without SUV39H rebinding}
When SUV39H does not rebind to its product, the degree of condensation dramatically drops (Fig.4B) following JMJD2 increase. As this system is monostable, the initial condensation should be recovered sooner or later with the limited number of actors taken into account here, but a too long period of decondensation in the cell can in practice offer the opportunity to other mechanisms to modify the cellular state by derepressing unwanted gene expression.

\end{multicols}
\begin{center}
\includegraphics[width=13cm]{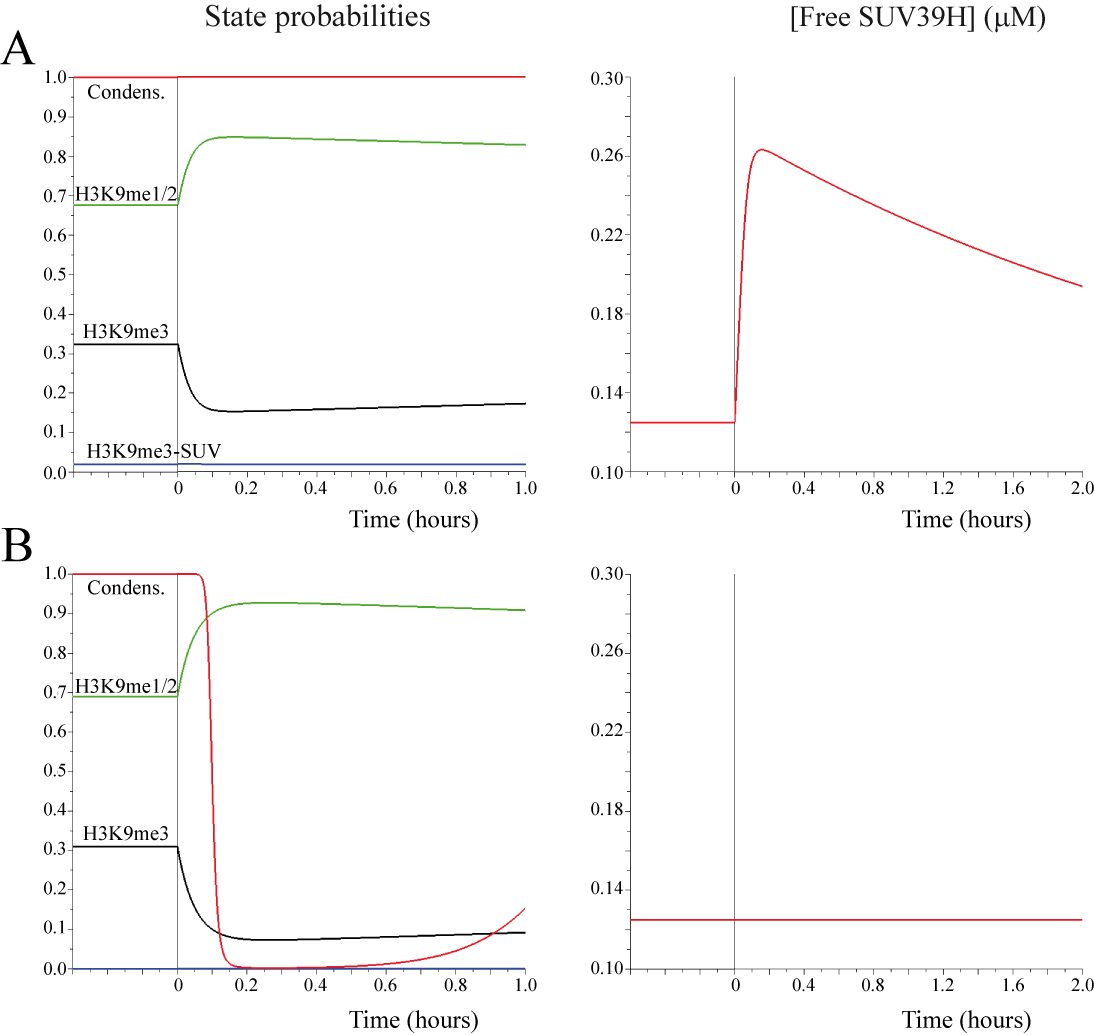} \\
\end{center}
\begin{small} \textbf{Figure 4}. Influence of the rebinding activity of SUV39H on nonlinear condensation visualized using the realistic parameters listed in SI-B. Identical bursts of JMJD2 are applied at time 0 in (\textbf{A}) with SUV39H rebinding to H3K9me3 and in (\textbf{B}) without rebinding ($ k_{3}=0 $) while keeping constant the other parameters. \end{small}\\ 
\begin{multicols}{2}

\subsection{Possible additional effects}

By itself, the phenomenon of enzyme rebinding is sufficient to explain the resistance of heterochromatin to stochastic pulses of demethylation; but additional parameters could complete this mechanism, such as the slow exchanges of SUV39H between heterochromatin niches and euchromatin located in the rest of the nucleosol, which can confine free SUV39H1 in heterochromatin and prevent euchromatin modification.

\subsubsection{Confinement of the SUV39H bursts in chromocenters}
This possibility is based on the high receptor density which can restrict the exchanges between chromocenters and the surrounding nucleosol. Homogeneous chromatin domains can form functional "niches" in which molecular movements are strongly slowered. This has been postulated for small metabolites \cite{Sassone} and is naturally expected for bulky enzymes and enzyme complexes. Slow protein diffusion has been revealed by fluorescence microscopy techniques and attributed to molecular crowding in heterochromatin areas \cite{Bancaud}, where phenomena like hopping between clustered binding sites, cage effects and collisions with obstacles, are supposed to be much more frequent \cite{Erdel2}. But the consequence of this situation on molecular binding is far from consensual. Slow diffusion is expected to decrease the frequency of encounter complexes, but molecular crowding is also expected to favour the binding of the encounter complexes, making difficult to predict the global result of these opposite influences \cite{Ellis}. In fact, dense obstacles can slower diffusion but are unable by themselves to create an enrichment of the free ligand in a particular compartment at equilibrium. However, high local concentration can be obtained transiently through a bottleneck effect. If the translocation rate between chromatin niches and the general nucleosol is low enough, transient rises of free SUV39H are obtained in the niche following a sudden release of SUV39H from H3K9me3 sites. For a minimal model in which the phenomenon of hopping is restricted to SUV39H, the system previously described should be completed as follows. We define $ S^{G} $ as the general concentrations of SUV39H in the nucleus, $ S $ its local concentrations in the chromocenters and $ k_{9} $ the symmetric translocation rate between the general and chromocenter compartments. The association rates used in the model takes crowding, supposed to be constant, into account. Eq.(3d) should be replaced by two equations. 

\begin{subequations} \label{E:gp}
\begin{equation} \dot{S}= N(k_{4}z-k_{3}Sy)-k_{9}(S-S^{G})\end{equation} \label{E:gp1}
\begin{equation} \dot{S^{G}}= k_{5}-k_{6}S^{G}+k_{9}(S-S^{G}) \end{equation} \label{E:gp2}
\end{subequations} 
and the initial conditions take the new values
\begin{subequations} \label{E:gp}
\begin{equation}  x_{0}= k_{2}k_{4}k_{6}^{2}k_{7}/D
\end{equation} \label{E:gp1}
\begin{equation}  y_{0}=k_{1}k_{4}k_{5}k_{6}k_{8}/D \end{equation} \label{E:gp2}
\begin{equation} z_{0}= k_{1}k_{3}k_{5}^{2}k_{8}/D \end{equation} \label{E:gp3}
\end{subequations}

where \textit{D} is the sum of the numerators.\\

Given the large amount of enzymes trapped in heterochromatin, the slow export rate of enzymes from chromatin niches to the general nucleosol creates a bottleneck effect confining the outbursts of free SUV39H (Fig.5), thus preventing accidental methylation of euchromatin. The lowest is the (unknown) rate constant $ k_{9} $, the strongest is this effect.

\end{multicols}
\begin{center}
\includegraphics[width=13cm]{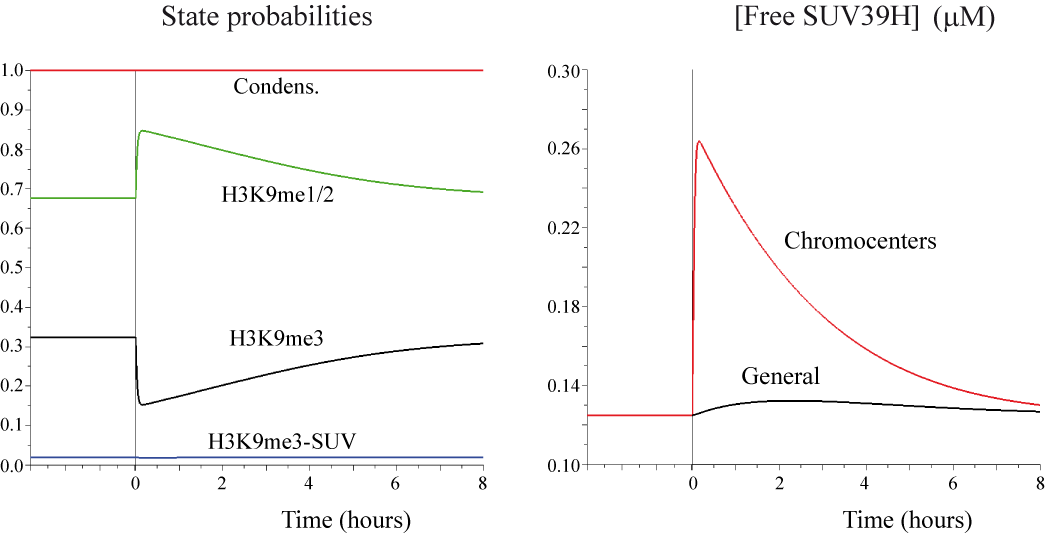} \\
\end{center}
\begin{small} \textbf{Figure 5}. Following a burst of JMJD2 at time 0, the reactive burst of free SUV39H is expected to remain confined in the chromocenters through a bottleneck effect in case of low exchange rate between the chromocenters and the nucleosol. Such a mechanism would be an additional advantage of heterochromatin segregation, restricting methylation to the chromocenters and minimizing the risk of H3K9 methylation of euchromatin. \end{small}\\ 
\begin{multicols}{2}

\subsubsection{H3K9me3-HP1-SUV39H ternary complexes}

The involvement of HP1 in the binding of SUV39H heterochromatin seems difficult to establish experimentally. It is supported by certain studies but not by others (Section 2.3). A recent unifying study proposed that SUV39H needs HP1 at low concentration but becomes self-sufficient at high concentration \cite{Muramatsu}. A participation of HP1 in the recruitment of SUV39H would be interesting during the establishment of heterochromatin as it would generate a complex cooperative behavior. The compaction of H3K9me3-rich chromosome regions can be driven by the simple bridging action of HP1$ \alpha /\beta $ following the minimal model of compaction of \cite{Johnson}, and could in turn intensify the recruitment of SUV39H, further increasing the density of H3K9me3, thereby recruiting more HP1 bridges, in a circular self-stabilizing circuit similar to the positive feedback loop of \cite{Brackley}, accentuating the spatial exclusion between H3K9me3 and H3K9ac-rich regions.

\section{Comparison with previous models}

\subsection{Positive vs negative feedbacks}
The earlier models of recruitment of chromatin-modifying enzymes mediated by their own products, were positive feedbacks \cite{Dodd,David-Rus,Sneppen}, whose ultrasensitivity and capacity to generate bistable states are well established. Positive feedbacks are particularly appropriate for the establishment, replication copying and mitotic memory of heterochromatin, while negative feedbacks rather ensure the stability of established marks, for instance in a life-long manner in post-mitotic cells. Positive and negative feedbacks can of course coexist in real systems. Precisely, the present model combines a positive feedback involved in the establishment of H3K9me3 heterochromatin through the mutual enforcement of H3K9me3 and SUV39H concentrations, and a negative feedback in case of demethylation by a reactive burst of soluble SUV39H. The positive feedback has two main effects illustrated in the simulation shown in Fig.6. The increase of H3K9me3 is slower because of the enzymatic inactivity of SUV39H trapped on H3K9me3, but this mechanism allows the generation of a stock of heterochromatin-bound SUV39H consistent with the potent accumulation of insoluble SUV39H1 detected in Fig.2. The slower activity of SUV39H imposed by its rebinding is related to the classical treatment of sequestration in enzymology:

\begin{equation} v= k_{cat}[SUV]_{tot}\dfrac{\dfrac{Ny}{K_{M}}}{1+\dfrac{Ny}{K_{M}}+\dfrac{Nz}{K_{d}}} \end{equation}

with $ K_{M} =k_{cat}/k_{1} $ and $ K_{d} =k_{3}/k_{4} $ (Fig.3).\\
\newline
But this decrease of the net enzyme activity is compensated by its accumulation which allows, by negative feedback, to robustly stabilize heterochromatin by absorbing stochastic bursts of demethylation, with a strength transiently proportional to the intensity of demethylation.

\subsection{Comparison with the model of nucleation and looping}

The most recent and comprehensive model of pericentric heterochromatin formation is centered on the most immobile fraction of SUV39H anchored in chromocenters \cite{Rippe}. It is postulated in the mechanism of nucleation and looping that fixed SUV39H methylates H3K9 by chromatin looping. This view was suggested by an experiment of anchorage of SUV39H in the nuclear lamina which expectedly caused an enrichment of H3K9me3 in the perinuclear area \cite{Rippe}. The alternative mechanism described here does not exclude this possibility, but is also compatible with other modes of chromocenter formation without anchorage, for example through the progressive merging of H3K9me3 regions, as described in \cite{Tessadori}. The cooperative mechanism of chromatin condensation mediated by HP1 \cite{Canzio,Teif,Mulligan} predicts a H3K9me3 threshold for chromocenter formation. The present model can explain both the formation of chromocenters and their subsequent maintenance, two aspects recently shown separable \cite{Ragunathan}.\\
The role of SUV39H rebinding to H3K9me3, which is central in the present study, has been minimized in \cite{Rippe} by arguing that the level of H3K9me3 is only slightly lower outside the chromocenters compared to chromocenters. The immunostaining of H3K9me3 and H3K9ac shown in Fig.S1A are clearly different in and out of the chromocenters. In addition, the global level of H3K9me3 could not be the only parameter intervening in chromocenters. In particular, the distribution of H3K9me3 along the chromosomes is different in chromosome regions containing repetitive and nonrepetitive DNA. The homogeneous repetitive organization of satellite DNA is prone to form large self-stabilized domains, whereas scattered H3K9me3 islands could be unable to create diffusion niches. Accordingly, long arrays of tandem repeats have long been shown efficient to create silenced heterochromatin, even if they are unrelated to natural satellite DNA \cite{Henikoff,Pecinka}. The spontaneous coalescence of heterochromatic structures sharing common interaction partners can be a spontaneous physical process which does not require the assistance of additional mechanisms \cite{Brackley,Hamada}, suggesting that the tandem repeat organization of satellite DNA is crucial for monotonous heterochromatin.

\end{multicols}
\begin{center}
\includegraphics[width=12cm]{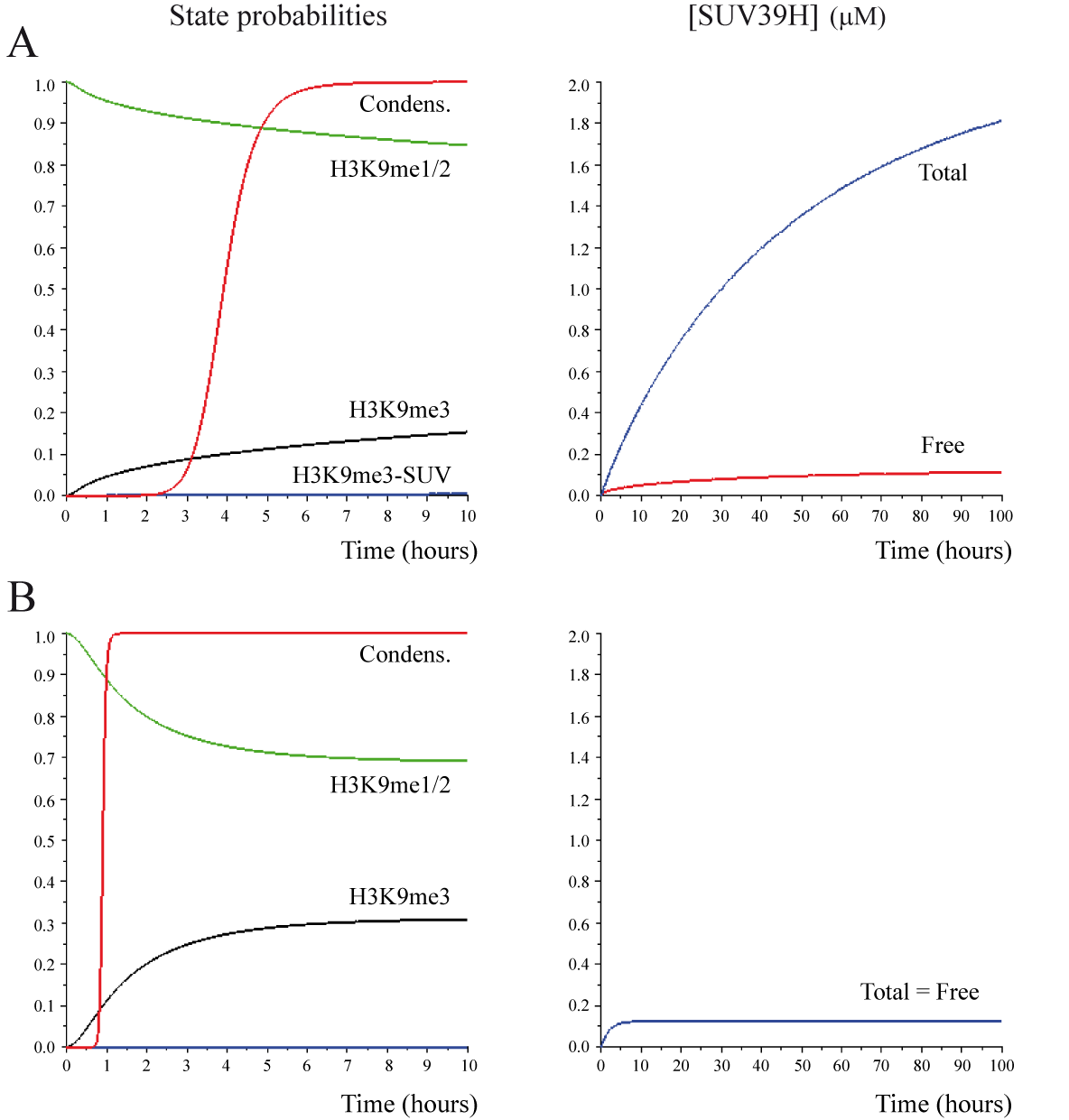} \\
\end{center}
\begin{small} \textbf{Figure 6}. Comparative accumulation of SUV39H and H3K9me3 with (\textbf{A}) and without (\textbf{B}) rebinding. (\textbf{A}) The mutual enforcement of H3K9me3 and SUV39H leads to a slow but extensive accumulation of total SUV39H. (\textbf{B}) In the purely theoretical hypothesis of an absence of rebinding of SUV39H to H3K9me3, the accumulation of H3K9me3 and condensation would be faster but would not generate a stock of SUV39H. These simulations use the same set of equations Eqs.(1)-(4) and the same parameters (SI-B) as previously, but with no initial H3K9me3. The total concentration of SUV39H is given by $ S+Nz $. \end{small}\\ 
\begin{multicols}{2}

\section{Conclusions}

The loss of heterochromatin is critical both in cancer and aging \cite{Tsurumi,Zhang}. The simple model proposed here can greatly contribute to the remarkable stability of H3K9me3 heterochromatin, securing the maintenance of differentiated cellular states and increasing the lifespan of adult vertebrates. The sequestration of SUV39H on H3K9me3 heterochomatin can regulate its local concentration by two means, possibly overlapping: (\textbf{i}) by saving it from the particular unstability of chromatin modifying enzymes \cite{Selbach} (which is particularly clear for SUV39H \cite{Khanal,Park,Bosch}) and (\textbf{ii}) by keeping it in chromatin niches. In turn, the high local concentration of SUV39H is expected to increase that of H3K9me3. These mechanisms are less demanding in term of biochemical conditions than the positive feedback models of heterochromatin spreading, mediated by subtle allosteric changes and by simultaneous physical contacts between several nucleosomes and modifying machineries. In the present model, the trapping of SUV39H on heterochromatin should not be conceived as a shield because (i) it is bound to a minor fraction of nucleosomes, and (ii) a shield requires immobile molecules to forbid the action of JMJD2. Instead, the binding of SUV39H to heterochromatin is a mean to increase H3K9me3 and create a latent source of free enzyme while maintaining constant its synthesis. This mechanism of "resistance to change" would authorize cellular reprogramming only upon sustained, nonrandom, upregulation of demethylases in the cell (Fig.7). High local concentration is recognized as a fundamental strategy of life \cite{Oehler}, but the concept of local concentration is often corrupted by a confusion between the concentration of free and bound molecules. The persistent sequestration of an enzyme with clustered receptors inevitably increases its local concentration, but without any functional consequence on the free enzyme, so that the modification of new substrates still requires the import of more enzymes. By itself, the phenomenon of enzyme trapping corresponds to a functional inactivation of the enzyme by sequestration, forbidding the modification of other substrates. But these enzymes can be massively de-sequestrated in case of bursts of JMJD2, for example during intermittent hypoxia, if not too prolonged. Considering the number of reported cases of protein sequestration in insoluble structures in the cell, this model could be generalizable to a variety of contexts, but the impressive arrays of nucleosomes in the nucleus appear ideally suited for the evolutionary selection of such a mechanism. H3K9me3 heterochromatin maintenance can also concern certain genes prepared for long-term transcriptional repression which are relocated in chromocenters \cite{Hubner,Guasconi}, where they can benefit from the higher local concentration of SUV39H. This mechanism is economic in that it necessitates a minimal number of ingredients to work. For instance it does not require a specific machinery of protein degradation, but simply the escape from normal turnovers of soluble protein by insolubilization on chromatin. 

\begin{center}
\includegraphics[width=7cm]{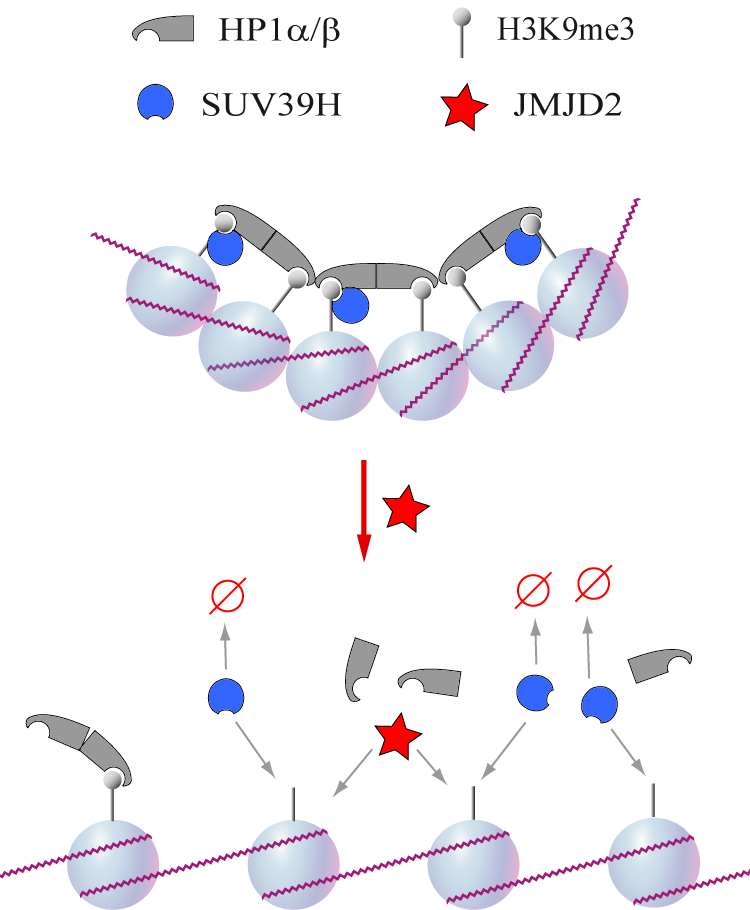} \\
\end{center}
\begin{small} \textbf{Figure 7}. Simplified scheme of the contribution of heterochromatin-bound SUV39H to the resistance of H3K9me3. The local accumulation of SUV39H can serve as a source of free SUV39H which can counteract demethylation unless the input of demethylating enzyme is strong and sustained enough, opening the way to reprogramming \cite{Antony}. The kinetic competition between the degradation ($ \O  $) and activity of free SUV39H allows to "sense" the duration of the demethylation phase. \end{small}\\ 

The resistance to degradation of insoluble proteins has long been illustrated by the numerous cases of natural and pathological aggregative proteins whose cellular content increases without modifying their expression level \cite{fibrils}.\\ 
As for near all proteins, SUV39H activity is subject to the whole panoply of regulatory mechanisms of molecular biology (which forbids exhaustive modeling), including synthesis \cite{Zheng}, degradation \cite{Khanal,Bosch}, alternative splicing, protein methylation \cite{Vaquero,Bosch}, phosphorylation \cite{Park}, cis-trans-isomerization \cite{Khanal} etc, which are likely to play refined regulatory roles in dividing cells with cycles of chromatin condensation/decondensation. But owing to its simplicity, the minimal model presented here could be an important facet of the simple maintenance of heterochromatin in differentiated cells. This mechanism is consistent with experimental observations and explains how dynamic epigenetic marks can ensure the long-term persistence of chromatin states, in apparent contradiction with the results of FRAP analyses showing that SUV39H interactions with chromatin are not static but dynamic over time scales below the hour.

\section{Acknowledgements} This work is supported by the Ligue inter-r\'egionale de lutte contre le cancer.\\

\textbf{Supplemental information} includes detailed material and methods, the set of parameter values used for the simulations and microscopy images showing the actors of this study.

\begin{small}

\end{small}
\end{multicols}

\newpage
\begin{center}
\huge{\textbf{Supplemental Information}}
\newline
\end{center}
\begin{multicols}{2}
\appendix
\setcounter{equation}{0}  
\numberwithin{equation}{section}

\section{Materials and Methods}

\subsection{Cell lines and plasmid transfection}

MCF7, MDA-MB-231, HepG2, Hela and 3T3 cells were routinely cultured in DMEM (GIBCO) supplemented with 10\% FBS (biowest) and antibiotics (Invitrogen) at 37°C in 5\% CO2. MCF7 cells stably expressing empty construct or deleted form of MKL1 (MKL1-$ \Delta $N200) construct were established using the T-Rex subclones system \cite{Flouriot}. MKL1-$ \Delta $N200 expression was performed by treatment with 1 $ \mu $g/ml tetracyclin, in DMEM supplemented with 2\% of FBS 48h prior to the experiment. For plasmids transfection, 0.5 $ \mu $g of DNA was used per well in 24-wells plate, using JetPEI according to manufacturer's instructions. The GFP-tagged SUV39H1 plasmids were kindly provided by Dr Krouwels. GFP-HP1alpha was a gift from Tom Misteli (Addgene plasmid \# 17652). JMJD2-GFP plasmids was kindly provided by Dr Nicholas Shukeir. Mutagenesis was performed using Quickchange primer Design from Agilent. 

\subsection{RNA extraction and qPCR}

RNA extractions were performed using RNeasy kit (Qiagen). Retrotranscription was performed on 1 $ \mu $g of RNA with random primer using MMLV reverse transcriptase. Quantitative RT-PCRs were performed using the iQ$ ^{TM} $ SYBR-Green supermix from BioRad (Bio-Rad, Hercules, CA, USA). The primer sequences used for SUV39H1 qPCR were: forward  ACCTGGTTGAGGGTGATGC and reverse CAGAAGGCCAAGCAGAGG.

\subsection{Immunoblotting}

For endogenous SUV39H1 immunoblotting, cells were lysed by sonication in 2X Laemmli buffer. Transfected cells expressing HA-tagged SUV39H1 or JMJD2C cells were lysed for 1 h on ice in lysis buffer (150 mM NaCl, 50 mM Tris-HCl, 1\% (v/v) Nonidet P-40, pH 8.0) supplemented with Complete protease inhibitor cocktail (Roche Diagnostics S.A., Meylan, France). Soluble and insoluble proteins were separated after a 30-min centrifugation at 14 000 g. The various protein extracts were loaded on SDS-PAGE and electrotransferred to a nitrocellulose membrane (Millipore). The blots were probed with primary antibodies: H3 (ab 1791, Abcam), HA (Y11, Santa Cruz biotechnology), SUV39H1 (D11B6, Cell signaling)

\subsection{Immunofluorescence}

Cells were plated on cover slides in 24 well plates. After treatment, cells were fixed with PBS containing 4\% paraformaldehyde for 10 min, and then permeabilized in PBS containing 0.3\% Triton X-100 for 10 min. The cells were incubated overnight at 4°C with primary antibodies (1/1000) in PBS/3\% FCS. Incubation with secondary antibody was performed for 2 h at room temperature. Images were obtained with an Imager Z1 ApoTome AxioCam (Zeiss) microscope. The comparisons between cells and conditions were made on average signals per surface unit (kept identical between the different cells) and quantified from images obtained with identical time exposures. Immunofluorescence was scored for at least 20 cells by image themselves obtained using ImageJ from different experiments. Primary antibodies used are H3K9Ac (ab12179), H3K9me3 (ab8898) and JMJD2C (ab27532) from Abcam, SUV39H1 (D11B6) from Cell Signaling, FLAG-tag (F3165) from Sigma, HA-tag from Santa Cruz (sc805).

\newpage

\section{Values used for the simulations shown in figures}

\end{multicols}

\begin{table}[h!]
\centering
\caption{Values used for the simulations shown in figures.}
\label{tab:1}
\begin{tabular}{l c c}
& & \\
\hline
& & \\
Parameter & Value & Reference  \\
\hline
& & \\
 Nucleosome concentration in the nucleus & 100 $ \mu $M &  Calculated$ ^{a} $ \\
& & \\
($ k_{1} $) SUV39H catalytic efficiency & $ 0.18/ \mu $M.min  & \cite{Jeltsch} $ ^{b} $\\
& & \\
($ k_{2} $) JMJD2 catalytic efficiency & $ 0.04/ \mu $M.min  & \cite{Hillringhaus,Krishnan} $ ^{b}, ^{c} $ \\
& & \\
($ k_{3} $) $ k_{on} $ of SUV39H1 & 1.05 /$ \mu $M.min & \cite{Rippe}  \\
& & \\
($ k_{4} $) $ k_{off} $ of SUV39H1 & 2 /min  & \cite{Rippe} $ ^{d} $ \\
& & \\
($ k_{6} $) SUV39H degradation rate  & 0.008 /min  &  \cite{Bosch} \\
& & \\
($ k_{8} $) JMJD2 degradation rate  & 0.008 /min  & \cite{Mallette} $ ^{e} $ \\
& & \\
($ k_{5} $) SUV39H synthesis rate  & 0.001 $ \mu $M/min  & Deduced from its concentration$ ^{f} $ \\
& & \\
($ k_{7} $) JMJD2 synthesis rate  & 0.01 $ \mu $M/min  & To give 30\% H3K9me3 \cite{Fodor} $ ^{g} $ \\
& & \\
($ k_{9} $) Chromocenter-nucleosol exchange rate   & 10$ ^{-3} $/min  & Arbitrary$ ^{h} $  \\
& & \\
($ K $/HP1) Condensation threshold constant  & 0.1 & For convenience \\
& & \\
($ \bar{n} $) Hill coefficient of cooperative condensation  & 20  & For convenience \\
& & \\
\hline
\end{tabular}
\end{table}
\begin{multicols}{2}

\subsection*{Values estimation}
\begin{itemize}
\item $ ^{a} $ A nucleosome every 200 bp along a 2 $ \times \ 3 \times 10^{9} $ bp-long diploid DNA in a spherical nucleus of 500 $ \mu $m$ ^{3} $, yields 100 $ \mu $M nucleosomes.

\item $ ^{b} $ For a given nucleosome, the pseudo first order enzyme binding rate is $ k_{a}E $ (where $ k_{a} $ is the second-order association constant and $ E $ is the concentration of the free enzyme). This rate should be weighted by the probability that once bound, the enzyme catalyses the reaction (rate constant $ k_{c} $) instead of dissociating from the nucleosome (rate constant $ k_{d} $). The probability of this event is $ k_{c}/(k_{c}+k_{d}) $ Hence, for a given nucleosome, the resulting global transformation rate is

$$ kE=  k_{a}E \dfrac{k_{c}}{k_{c}+k_{d}} = \dfrac{k_{c}}{K_{M}}E $$
This linear approximation does not take enzyme sequestration into account, but the error is moderate for enzymes with relatively large $ K_{M} $ values (around 30-40 $ \mu $M).

\item $ ^{c} $ The catalytic efficiency of JMJD2 is set from the measurements for JMJD2C ($ 0.036/ \mu $M.min) \cite{Hillringhaus}, for JMJD2D ($ 0.045/ \mu $M.min) and JMJD2A ($ 0.031/ \mu $M.min) \cite{Krishnan}.

\item $ ^{d} $ The on and off rates for the interaction between SUV39H1 and H3K9me3 are derived from the fluorescence microscopy analyses of \cite{Rippe}. The off rate corresponds to the moderately mobile fraction (class III) of SUV39H defined in \cite{Rippe}.

\item $ ^{e} $ Half-lifes of 90 \cite{Mallette}, 120 \cite{Harper} and 180  minutes \cite{Capucine}, have been found for JMJD2A and 60 minutes for JMJD2B \cite{Ipenberg}, so a median value of 90 min is selected for the simulation. 

\item $ ^{f} $ Given the degradation rate of SUV39H, its synthesis rate is determined to be compatible with the steady state (synthesis/degradation) concentration of dimers of 0.05$ \mu $M  \cite{Rippe}. 

\item $ ^{g} $ Given the previous values, the collective synthesis rate of JMJD2 enzymes is adjusted to yield a steady state level of trimethylated H3K9 of 30\% according to \cite{Fodor}.

\item $ ^{h} $ This value is currently unknown. The bottleneck effect described in the article begins at moderate level for relatively low rate constants and increases as this constant decreases.

\end{itemize}

\section{Visualization of the actors of the system}

The different actors of the model proposed here are presented in mouse 3T3 cells, in Figs.S1 and S2 summarizing results scattered in the literature. The coalescence of H3K9me3-rich regions superposes well with areas intensely colorable with DAPI and identified as chromocenters. HP1$ \alpha $ and SUV39H1 are enriched in these regions but the density of acetylated H3K9ac is reduced (Fig.S1A). The catalytic activity of SUV39H1 is not necessary for this enrichment, suggesting that this is a rebinding phenomenon (Fig.S1B). The JMJD2B demethylase is not specifically enriched in the chromocenters. Only when catalytically active, JMJD2B prevents the enrichment of SUV39H1 in chromocenters (Fig.S2).

\begin{small}

\end{small}

\end{multicols}
\begin{center}
\includegraphics[width=16cm]{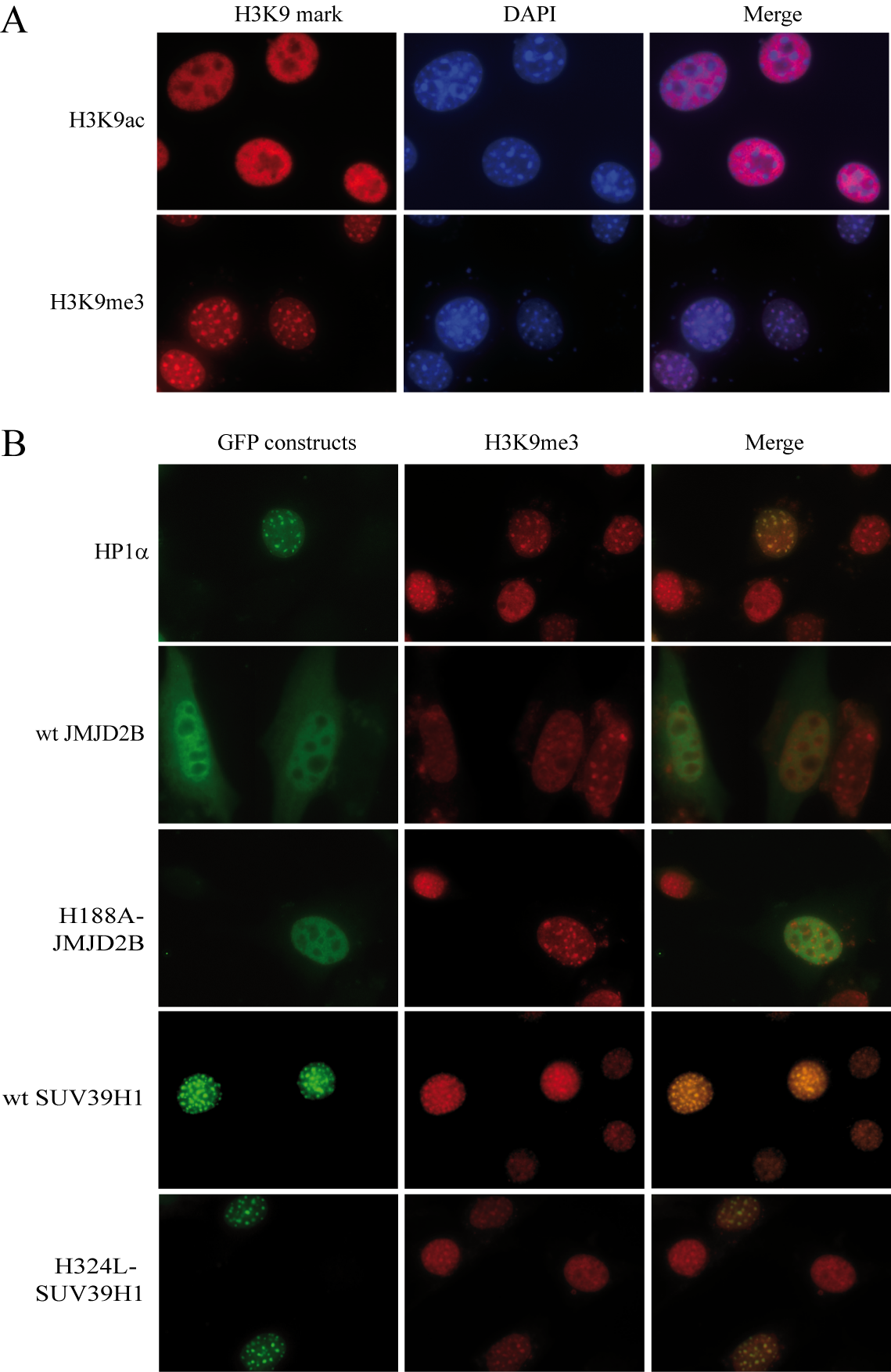} \\
\end{center}
\begin{small} \textbf{Figure S1}. (\textbf{A}) Spatial segregation of trimethylated and acetylated H3K9me3. H3K9me3 is clearly enriched in the chromocenters which are intensely stained with DAPI, while conversely, acetylated H3K9 appears excluded from these areas.  (\textbf{B}) Localization of GFP-tagged constructs relatively to H3K9me3. HP1$ \alpha $, wild type SUV39H1 and the catalytically inactive mutant (H324L) of SUV39H1 are similarly enriched in H3K9me3-dense regions. H3K9me3 labeling is strikingly intensified in cell transfected with wt SUV39H1 compared to surrounding non-transfected cells and to the cells transfected with the catalytic mutant of SUV39H1. Note that overexpression of JMJD2B, but not of its catalytically inactive mutant (H188A), tends to soften the H3K9me3 dots.
\end{small}\\ 
\newline 
\newline 
\newline 
\newline 
\begin{center}
\includegraphics[width=10cm]{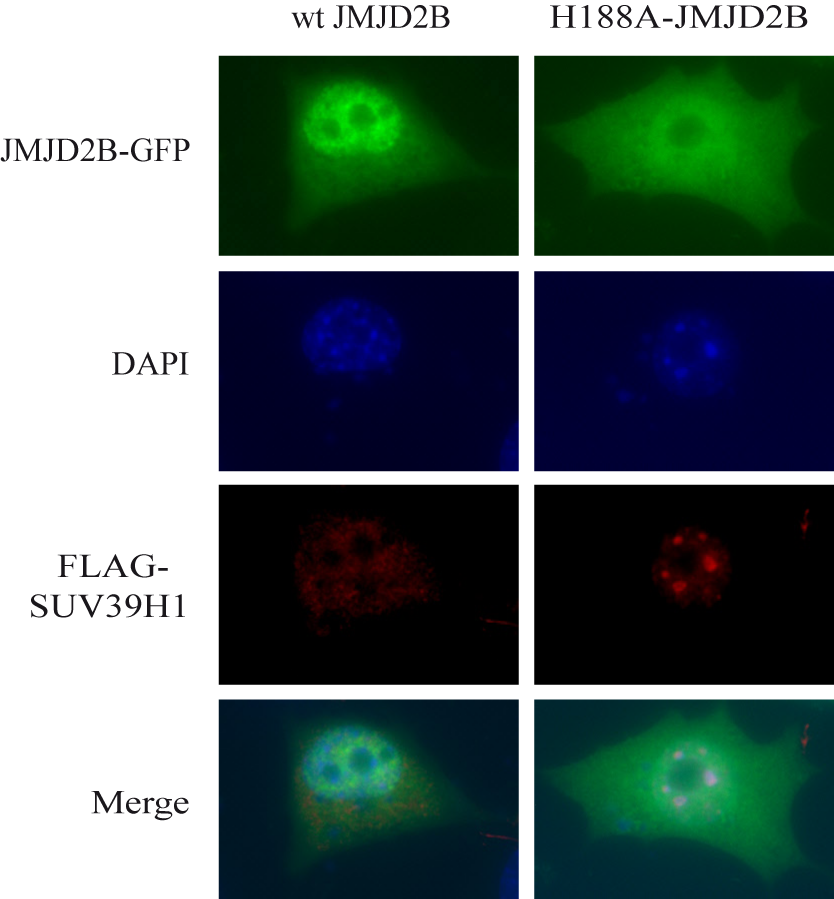} \\
\end{center}
\begin{small} \textbf{Figure S2}. Altered spatial distribution of FLAG-SUV39H1 in cells co-transfected with a catalytically active version of JMJD2B-GFP. The strong accumulation of exogenous FLAG-SUV39H1 in chromocenters is impeded by the catalytically active JMJD2B-GFP but is not pertubated by its catalytic mutant. \end{small}\\ 

\end{document}